\documentclass[sigconf]{acmart}
\renewcommand\footnotetextcopyrightpermission[1]{} 
\usepackage{fancyhdr}
\usepackage{booktabs} 
\usepackage{listings}
\usepackage{amssymb}
\usepackage{enumitem}
\usepackage{multirow}
\fancypagestyle{plain}{%
\fancyhf{} 
\fancyfoot[C]{\large\thepage} 

}
\DeclareMathOperator*{\argmin}{arg\,min}

\begin{document}
\title{NECTAR: Non-Interactive Smart Contract Protocol using Blockchain Technology}

\author{Alexandra Covaci}
\affiliation{%
  \institution{nChain}
  \city{London}
  \country{United Kingdom}
}
\email{alexandra@ncrypt.com}

\author{Simone Madeo}
\orcid{0000-0002-7186-8240}
\affiliation{%
  \institution{nChain}
  \city{London}
  \country{United Kingdom}
}
\email{simone@ncrypt.com}

\author{Patrick Motylinski}
\orcid{}
\affiliation{%
  \institution{nChain}
  \city{London}
  \country{United Kingdom}
}
\email{patrick@ncrypt.com}

\author{St\'ephane Vincent}
\orcid{}
\affiliation{%
  \institution{nChain}
  \city{London}
  \country{United Kingdom}
}
\email{vincent@ncrypt.com}

\renewcommand{\shortauthors}{A. Covaci, S. Madeo, P. Motylinski, and S. Vincent}

\begin{abstract}
Blockchain-driven technologies are considered disruptive because of the availability of dis-intermediated, censorship-resistant and tamper-proof digital platforms of distributed trust. Among these technologies, smart contract platforms have the potential to take over functions usually done by intermediaries like banks, escrow or legal services.
In this paper, we introduce a novel protocol aiming to execute smart contracts as part of a blockchain transaction validation. We enable extensions in the execution of smart contracts while guaranteeing their privacy, correctness and verifiability. Man-in-the-middle attacks are prevented, since no communication between participants is requested, and contract validations do not imply the re-execution of the code by all the nodes in the network. However, proofs of correct execution are stored on the blockchain and can be verified by multiple parties.
Our solution is based on programming tools which optimize the time execution and the required memory while preserving the embedded functionality.
\end{abstract}

\keywords{Smart contract, blockchain, Bitcoin, verifiable computation, non-interactive protocol.}

\maketitle

\section{Introduction}
Blockchain, the core technology of cryptocurrencies, is generating significant interest across a wide range of industries, promising to support the redesign of interactions in business, politics and society at large.
A permission-less \textit{blockchain} network can be seen as a global, public log that records transactions between cryptocurrency clients in a decentralized manner, with internal consistency maintained through a distributed consensus mechanism~\cite{Dhillon2017}.
The state of the world in the Bitcoin blockchain is represented by a series of messages called transactions~\cite{Bonneau2015}. Bitcoin transactions have locking and unlocking mechanisms based on a scripting language which is \textit{primitive recursive}, thus lacking expressive power.
Transactions are recorded in blocks, each block being linked-back to the previous one through its hash value. Hence, blockchain security is established by this chain of cryptographic hashes solved by a loosely-organized network of participants called miners. 

Blockchain-driven technologies are considered disruptive because of the availability of dis-intermediated, censorship-resistant and tamper-proof digital platforms of distributed trust~\cite{mattila2016blockchain}.
Among the potential uses and innovations for blockchains, smart contract platforms have the potential to take over functions usually done by intermediaries like banks, escrow or legal services. However, for their broad adoption, there is still a need for developing easily verifiable protocols that respect the confidentiality and privacy of data, and this is a gap addressed by the current publication.

In this paper, we introduce \textbf{NECTAR}, i.e.
\textbf{N}on int\textbf{E}ractive smart \textbf{C}on\textbf{T}r\textbf{A}ct p\textbf{R}otocol, aiming to execute smart contracts as part of a blockchain transaction validation\footnote{NECTAR's technologies are the subject of the following UK patent applications:
\textbf{1718505.9} (9/11/17), \textbf{1719998.5} (30/11/17), \textbf{1720768.9} (13/12/17),
\textbf{1801753.3} (2/1/18).
}. The protocol is \textit{non-interactive}, i.e. no direct communication between the parties is required during the verification stage.
We concentrate on investigating the use of advanced cryptographic techniques to enhance and expand the blockchain capabilities in the context of smart contracts. Our proposed solution lies within the intersection of cryptography and formal verification, enabling extensions and innovations in the execution of smart contracts while guaranteeing their privacy, correctness and verifiability. Its principal benefits are (i) man-in-the-middle attacks are prevented, since no communication between participants is requested; (ii) malicious nodes cannot tamper with the data due to the use of blockchain technologies; (iii) contract validations do not require the re-execution of the code by all the nodes in the network, like in Ethereum~\cite{Ethereum}.


Three different entities operate in our protocol: the \textit{client} creates the contract, the \textit{worker} evaluates its computation using a given set of input values and produces a Proof of Correctness (PoC), while the \textit{verifier} validates the contract by checking the PoC.
In building NECTAR, our main contributions can be summarized as follows:
\begin{itemize}[leftmargin=0.3in]
\item \textit{Minimal contract execution}: contract validations do not require code re-execution and computations are not replicated by every node in the network.
\item \textit{Practical formal verification}: proofs of correct execution of the smart contracts are stored on the blockchain and can be verified by multiple parties.
\item \textit{Outsourcing of the contract execution}: a worker produces a certificate of correctness that can convince untrusted parties of the validity of the contract.
\item \textit{Functional correctness}: during verification, the worker publishes a non-interactive proof as part of a transaction and a verifier may accept or reject it.
\item \textit{Reduced space and time complexity}: our solution is based on programming tools which optimize the time execution and the required memory while preserving the embedded functionality. 
\end{itemize}
The paper is structured as follows: a background on smart contracts, verifiable computation and algebraic tools used in our protocol is presented in Section~\ref{sec_bg}. The details of NECTAR are introduced in Section~\ref{sec_proto}, with specific focus on the compiling pipeline for the translation of a contract written in a high-level language to a suitable arithmetic representation. In Section~\ref{sec_block}, we illustrate the interaction with the blockchain using the Bitcoin \texttt{Script} language. Finally, conclusions are presented in Section~\ref{sec_conc}.

\section{Background}
\label{sec_bg}
This section focuses on introducing the reader to the main technologies that underpin NECTAR: the existing smart contract platforms and the required mathematical tools.

\subsection{Smart contracts} 
Although commonly known as the technology underpinning cryptocurrencies such as Bitcoin~\cite{Nakamoto2008}, blockchain applications have increasingly gone beyond digital currencies  ~\cite{Bonneau2015}. Blockchains can be regarded as computational engines for digitizing asset ownership, intellectual property and the execution of smart contracts. 

The term \textit{smart contract} is generally used to describe a computer protocol that automatically facilitates, executes and enforces a contract made between two or more counterparties, removing the need for contractual clauses and recourse to the law. The idea behind smart contracts dates back to the mid 1990s, when Nick Szabo predicted that the digital revolution would drastically change the way humans make contracts~\cite{Szabo1997}. The rules of a contract can be encoded in a program that is replicated and executed across blockchain nodes. Privacy-friendly and secure contracts encrypt information using a public key or a commitment scheme, while transaction validations enable the verification of the correct execution of the smart contract. This process can employ Zero-Knowledge (ZK) proofs, Succinct Non-interactive ARguments of Knowledge (SNARK) proofs or a combination of both. Finally, the results of the secure computation process are stored on the blockchain~\cite{Sanchez2017}.

\begin{table}
\caption{Smart contract technologies and available features.}
\begin{center}
\begin{tabular}{ l c c c }
\hline
\multirow{2}{*}{\small{Technology}} & \multirow{2}{*}{\small{Non-interactive}} & \multirow{2}{*}{\small{Trustless}} & \small{Minimal contract}\\
 &&& \small{execution}\\
\hline
Ethereum~\cite{Ethereum} & \checkmark
 & \checkmark
 &  \\
ZKCP~\cite{Sudoku, Maxwell2015} &  & \checkmark
 & \checkmark
 \\
Hawk~\cite{Hawk2016} & \checkmark
 &  & \checkmark
 \\
\hline
\end{tabular}
\end{center}
\label{ref_table}
\end{table}

Existing smart contract applications built on top of Bitcoin (e.g. lotteries~\cite{Bentov2014} or multi-party computation~\cite{Kumaresan2014}) experience difficulties interfacing with the Bitcoin scripting language~\cite{Hawk2016}. These served as motivation for the development of other blockchain scripting languages and platforms that are better suited to smart contracts. 
New emerging technologies (e.g. Ethereum, Counterparty~\cite{Counterparty}) accelerate the evolution of smart contracts by extending Bitcoin design through a rich  Turing-complete bytecode language. Ethereum has a flexible interface that enables a large variety of applications for smart contracts. Additionally, the latest 
Ethereum Metropolis
(Byzantium)  software upgrade made possible privacy advancements that enable zk-SNARK-infused contracts~\cite{Metropolis}.

However, not all the smart contract solutions are based on zero-knowledge protocols. Counterparty extends Bitcoin with advanced financial operations,
e.g. creation of virtual assets and  payment of dividends,
by embedding its metadata into Bitcoin transactions recognized and interpreted by Counterparty nodes. Stellar~\cite{Stellar} is an open source platform that enables the provision of affordable financial services to people who have never had access to them (with a focus on Africa). Stellar is governed by a consensus algorithm inspired by the federated Byzantine agreement \cite{mazieres2015stellar}, where a node agrees on a transaction if the nodes in its neighborhood agree as well, thus consuming less computing power compared to proof-of-work. Stellar's virtual currency is called lumens, but it also allows users to retain other assets, e.g. telephone minutes. Monax~\cite{Monax} is a proof-of-stake smart contract-enabled blockchain that allows users to create private blockchains and define dedicated authorization policies.

Zero-knowledge cryptography on the blockchain was pioneered by Zcash~\cite{hopwood2016zcash}, a global open payment network, which can be used as a method for verifying a ledger entry without revealing the identity of any parties. Hawk~\cite{Hawk2016} is a zero-knowledge based framework for building privacy-preserving smart contracts that provides both programmability and transaction privacy. Hawk includes a correct-by-con\-struc\-tion compiler for user-defined applications that allows any non-specialist programmer to write a program without implementing any cryptography. Trust assumptions introduced by Hawk-generated protocols rely on minimally trusted managers, who can see the user's inputs but cannot affect the correct execution of the contract.
Zero Knowledge Contingent Payment (ZKCP) is another protocol based on zero-knowledge techniques that allows fair exchange over the Bitcoin blockchain~\cite{Sudoku, Maxwell2015}. ZKCP relies on two processes: an atomic swap over the blockchain and an interactive zero-knowledge scheme, where communication between the parties is necessary, making the protocol susceptible to denial-of-service attacks.

In Table~\ref{ref_table}, we present the most important features required to build privacy-friendly and secure contracts, showing how the above-mentioned technologies fulfill them. We observe that none of the existing solutions ticks all the boxes that would offer an ideal minimal set of functionalities.

\subsection{Algebraic tools}
In the following sections, we use the term smart contract to describe general purpose computations that take place on a blockchain and are influenced by external events. 
In recent seminal works~\cite{Gennaro12,Parno16}, it was shown how to compactly encode computations as quadratic programs, in order to provide non-interactive, publicly verifiable computations. 
The basic definitions of arithmetic circuits, quadratic arithmetic programs and bilinear groups are now provided.


\subsubsection{Basic Notation}
We denote by $\mathbb{G}$ a group, and consider only groups that are cyclic and have prime order $r$. 
Group elements are denoted with calligraphic letters, such as $\mathcal{P}$ and $\mathcal{Q}$. 
Given a group $\mathbb{G}$, we say that $\mathcal{P}$ generates $\mathbb{G}$, i.e. $\mathbb{G}=\langle\mathcal{P}\rangle$, and use additive notation for group arithmetic. 
Hence, $\mathcal{P}+\mathcal{Q}$ denotes addition of the elements  $\mathcal{P}$ and $\mathcal{Q}$; $a\cdot\mathcal{P}$ denotes scalar multiplication of $\mathcal{P}$ by the scalar $a \in \mathbb{Z}$. 
We denote by $\mathbb{F}$ a field, and by $\mathbb{F}_p$ the field of order $p$. 
We consider only fields of prime order.
We denote by $\mathbb{F}_{p^k}$ the extension field of degree $k$ of $\mathbb{F}_p$, where $k$ is a non-zero integer. 
Let $E$ be an elliptic curve over the finite field $\mathbb{F}_p$ (resp. $\mathbb{F}_{p^k}$), we sometimes write $E(\mathbb{F}_p[r])$~(resp. $E(\mathbb{F}_{p^k}[r]))$ to denote the abelian subgroup of order $r$ of $E(\mathbb{F}_p)[r]$~(resp. $E(\mathbb{F}_{p^k})[r]$). 

\subsubsection{Modeling computations as Quadratic Arithmetic Programs}
\label{sec_qap}
An arithmetic circuit $C$ over a finite field $\mathbb{F}$ and a set of variables $\vec{x}=(x_1,...,x_k)$ consists of indegree 2 addition and multiplication gates and a set of wires between the gates. 
The wires carry values over $\mathbb{F}$. 
Every gate in $C$ of indegree 0 is labeled by either a variable from $\vec{x}$ or a field element from $\mathbb{F}$. 
We always regard an arithmetic circuit as computing a polynomial in $\mathbb{F}[x]$.
Unless otherwise stated, an arithmetic circuit $C$ has $d$ multiplication gates and $k$ wires.
The wires $(1,...,n)$ occupy inputs and outputs and the set $(n+1,...,k)$ represents the internal wires.

A quadratic arithmetic program (QAP) is a way of encoding arithmetic circuits, and some more general computations, over a field $\mathbb{F}$ of prime order $p$ given by a collection of polynomials over $\mathbb{F}$. 
For any function $f$ represented by an arithmetic circuit, we can easily construct a QAP that evaluates the function $f$. 
\begin{definition}\cite{Gennaro12} 
A QAP $Q$ over a field $\mathbb{F}$ is a tuple $Q$
\begin{equation}
Q=\Big({v_i(x)}_{i=1}^{k},~{w_i(x)}_{i=1}^{k},~{y_i(x)}_{i=1}^{k},~t(x)\Big),
\end{equation}
with $v_i(x),~w_i(x),~y_i(x)\in\mathbb{F}[x]$ polynomials of degree at most $d-1$. 
The polynomial $t(x)\in\mathbb{F}[x]$ is called a \textit{target polynomial} and has degree $d$. 
The size of the QAP is $k$ while its degree is $d$. 
We say that $Q$ evaluates a function $(a_{n+1},...,a_k)=f(a_1,...,a_n)$ if the tuple $(a_1,...,a_n)\in\mathbb{F}^{n}$ is a valid assignment of the inputs and outputs of $C$ and there exists a tuple $(a_{n+1},...,a_k)$ such that $t(x)$ divides 
$p(x)=\big(\sum_{i=1}^{k}a_iv_i(x)\big)\cdot\big(\sum_{i=1}^{k}a_iw_i(x)\big)-\big(\sum_{i=1}^{k}a_iy_i(x)\big)\in\mathbb{F}[x]$, i.e. there exists some polynomials $h(x)$ such that $h(x)\cdot t(x)=p(x)$.  
\end{definition}

The very basic intuition for building a QAP is to encode the input-output correctness for each gate in the polynomials $v_i(x)$,~$w_i(x)$ and~$y_i(x)$. 
For each multiplication gate $g$ this is done by first selecting an arbitrary value $r_g\in\mathbb{F}$ (i.e. a \textit{root}) and then, for every left wire $i$ going to gate $g$, one imposes $v_i(r_g)=1$. 
A similar process is done for polynomials $w_i(x)$ and $y_i(x)$ w.r.t. right input and output wires respectively. 
The target polynomial $t(x)$ is defined over the roots $r_g$: $t(x)=\sum_{i=1}^{d}(x-r_i)$.

\subsubsection{Bilinear Groups}
The protocol as described herein is based on bilinear pairing cryptography, i.e. the polynomials of a QAP are encoded into elements of groups. We assume two cyclic additive groups $\mathbb{G}_1$ and $\mathbb{G}_2$ of prime order $r$ with generators $\mathcal{P}$ and $\mathcal{Q}$ respectively together with map $e:\mathbb{G}_1\times\mathbb{G}_2\rightarrow\mathbb{G}_T$, where $\mathbb{G}_T$ is a multiplicative group of order $r$. 
The map $e$ is assumed to be bilinear, i.e. $\forall \mathcal{P}\in\mathbb{G}_1$, $\mathcal{Q}\in\mathbb{G}_2$, $a, b\in\mathbb{Z}_r$:
\begin{equation}
e(a\mathcal{P}, b\mathcal{Q})=e(\mathcal{P},\mathcal{Q})^{ab}. 
\end{equation}
The map $e$ is non-degenerate, i.e. for $\mathcal{P}\neq0_{\mathbb{G}_1}$ and $\mathcal{Q}\neq0_{\mathbb{G}_2}$, $e(\mathcal{P}, \mathcal{Q})=1_{\mathbb{G}_T}$; where $0_{\mathbb{G}_1}$ (resp. $0_{\mathbb{G}_2}$ and $1_{\mathbb{G}_T}$) is the neutral of the group $\mathbb{G}_1$ (resp. $\mathbb{G}_2$ and $\mathbb{G}_T$).
There are many ways to set up bilinear groups,
and our construction uses asymmetric bilinear groups where $\mathbb{G}_1\neq\mathbb{G}_2$.

The idea of \textit{encoding} consists in evaluating, for example, the polynomials $v_i(x), w_i(x)$ at a random element $s\in\mathbb{F}$ and mapping these elements to $v_i(s)\cdot\mathcal{P}$ in $\mathbb{G}_1$ and $w_i(s)\cdot\mathcal{Q}$ in $\mathbb{G}_2$. 
The calculations in the polynomial ring $\mathbb{F}[x]$ are translated into calculations in the exponent of $\mathbb{G}_T$ by $e\big(v_i(s)\cdot\mathcal{P},w_i(s)\cdot\mathcal{Q}\big)=e\big(\mathcal{P},\mathcal{Q}\big)^{v_i(s)w_i(s)}$.

\subsubsection{Proving Correctness of Computations}
For the worker to prove that an assignment $(a_{1}, \dots, a_n)$ on input/output wires is valid, it suffices to prove that there exists $(a_{n+1},...,a_k)$ corresponding to assignments on the internal wires such that $p(x)$ has roots $(r_1, \dots, r_d)$.
Each polynomial of the quadratic program, e.g. $v_i(x)\in\mathbb{F}[x]$, is mapped to an element $v_i(s)\cdot\mathcal{P}$ in a bilinear group, where $s$ is a secret value selected by the client, $\mathcal{P}$ is a generator of the group, and $\mathbb{F}$ is the field of discrete logarithms of $\mathcal{P}$. 
We refer to these elements as public parameters or a \textit{common reference string}. 
Thus, for a given input, the worker evaluates the circuit directly to obtain the output and the values of the internal circuit wires which correspond to the coefficients $a_i$. 
To oversimplify, the worker evaluates $v(s)\cdot\mathcal{P}=\sum_{i=1}^{k}a_i\cdot v_i(s)\cdot\mathcal{P}$ (resp. $w(s)\cdot\mathcal{Q}=\sum_{i=1}^{k}a_i\cdot w_i(s)\cdot\mathcal{Q}$ and $y(s)\cdot\mathcal{P}=\sum_{i=1}^{k}a_i\cdot y_i(s)\cdot\mathcal{P}$). 
The worker computes the polynomial $h(x)=p(x)/t(x)=\sum_{i=1}^{d}h_i\cdot x^i$ and constructs $h(s)\cdot\mathcal{Q}=\sum_{i=1}^{d}h_i\cdot(s^i\cdot\mathcal{Q})$. 
The elements $s^i\cdot\mathcal{Q}$ are public parameters. 
In this way, the worker is able to evaluate $h(s)\cdot\mathcal{Q}$ without learning the value $s$. 

In the last phase, the verifier uses bilinear pairings to check whether $t(x)$ divides $p(x)$:
\begin{equation}
e\big(v(s)\cdot\mathcal{P},w(s)\cdot\mathcal{Q}\big)~=~e\big(y(s)\cdot\mathcal{P},\mathcal{Q}\big)\cdot e\big(t(s)\cdot\mathcal{P},h(s)\cdot\mathcal{Q}\big).
\end{equation}

In existing QAP constructions, the worker does not know $s$, and hence cannot directly evaluate the polynomials $v_i(s), w_i(s), y_i(s)$ on each wire. 
In fact, security would be broken if the worker knew the value of the polynomials at $x=s$. 
The interested reader may also consult ~\cite{Gennaro12,Parno16,Schoenmakers2016}.

\begin{figure*}[t]
\includegraphics[width=\textwidth]{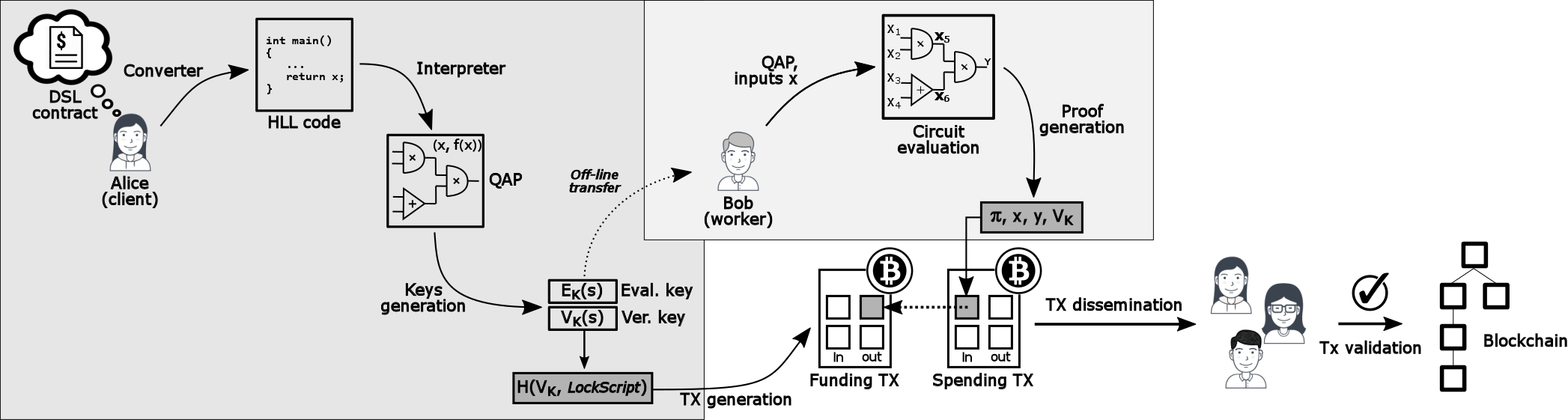}
\caption{
NECTAR is characterized by three phases: setup, evaluation and validation. During the \textit{setup phase}, a compiler/interpreter takes as input a contract source code, produces an arithmetic circuit and generates a quadratic program containing a set of polynomials that provides a complete description of the original circuit. The public parameters $(E_k, V_k)$ required by the worker and the verifiers are also generated. During the \textit{evaluation phase}, the worker evaluates the computation over a particular input $x$ and obtains the output $y$. Then, it uses the evaluation key $E_k$ to produce a proof-of-correctness $\pi$. The proof $\pi$ is stored on the blockchain and can be verified during the \textit{validation phase} by multiple parties without requiring the worker to interact with them. Every node can validate the payment transaction using the set $(\pi, x, y, V_k)$, thus validating the contract.}
\label{fig_main}
\end{figure*}

\section{NECTAR Protocol}
\label{sec_proto}
NECTAR allows non-specialist programmers to \textit{compose} smart contracts, \textit{outsource} the contract execution to untrusted parties and \textit{publicly verify} the correctness of the contract execution.
As illustrated in Figure~\ref{fig_main}, the protocol consists of three main phases. In the \textbf{setup phase}, contracts are written in a formal language with precise semantics. Contracts expressed in such a language have a mathematically precise meaning and can be manipulated by software. A compiler/interpreter takes as input the source code and produces an arithmetic circuit $C$ which consists of wires that carry values from a field $\mathbb{F}$ and connect to addition and multiplication gates. From the circuit $C$, the system first generates a quadratic program and then a set of public parameters that can be outsourced for execution to untrusted parties. During the \textbf{evaluation phase}, a worker evaluates the computation on a particular input $x$ and produces a PoC that is stored on the blockchain, based on the public parameters associated with the quadratic program. Finally, during the \textbf{validation phase}, the network nodes run a formal verification of the contract execution using the public parameters and the PoC.

Although more domain-specific languages (DSL) are required to implement a smart contract, e.g. Digital Asset Modeling Language (DAML) and Financial products Markup Language (FpML), as a first step we focus on a more generic language that provides a broader range of types, operators and constructs such as C. 
Our compilation pipeline is structured as follows. (i) The high-level C program containing the contract and the required external libraries are linked together to make the pre-processed contract. Pre-processor directives are evaluated. (ii) An intermediate-level language is a set of expressions of C-like operators, such as addition, multiplication, comparison, conditionals and logics.
(iii) The arithmetic circuit is built as symbols with wires connected to elementary arithmetic gates, e.g. addition and multiplication. (iv) The arithmetic circuit is optimized by exploiting mathematical and logic properties.
(v) Finally, the polynomials $(v, w, y)$ in the QAP are defined in terms of their evaluations at the roots of the arithmetic circuit, as explained in Section~\ref{sec_qap}.


\subsection{QAP-friendly curve}
When implementing pairing, one of the parameters on the curve is defined over the base field $\mathbb{F}_p$, and the other over $E(\mathbb{F}_{p^k})$.
A pairing-friendly elliptic curve needs to be added to the protocol in order to support the functionalities of NECTAR. 
Pairing-based cryptographic schemes require 
elliptic curves with a small embedding degree and a large-prime order subgroup that randomly generated elliptic curves are unlikely to have. 
It turns out that the elliptic curve group \texttt{secp256k1} used in Bitcoin has a fairly large embedding degree. 
We consider a class of curves with embedding degree $k=12$, but elliptic curve arithmetic over $\mathbb{F}_{p^{12}}$ can be computationally very expensive. However, by using an appropriate map, we can compress certain points in $E(\mathbb{F}_{p^{12}})$ to points in a twisted curve $\tilde{E}(\mathbb{F}_{p^{k/d}})$. 
We consider a sextic twist ($d=6$).  

We chose to build NECTAR by extending the Pinocchio C++ implementation~\cite{Parno16}. 
The original implementation of Pinocchio uses libraries that are not available for public use (Microsoft internals). 
The first step was to replace those libraries with available libraries that have similar characteristics. 
We use the GNU Multi-Precision (GMP) library~\cite{GMP06} for polynomial arithmetic along with the Pairing-Based Cryptography (PBC) library~\cite{Lynn06} for Tate-pairing over a Barreto-Naehrig curve~\cite{Deve07}.

\subsection{Smart contracts and circuit representation}
\label{sec_simone}


Our compiler is able to process a significant set of instructions natively supported by the  C language, such as static initializers, global functions and block-scoped variables, arrays and structs, pointers, function calls, conditionals and loops, arithmetic and bitwise Boolean operators.
We highlight that \textit{the target arithmetic circuits can only support expressions solvable at compile time}, therefore pointers and array dereferences must be known constants during the compilation. Dynamic references would require an impractical overhead on the circuit size due to the evaluation of conditional expressions. In a worst-case scenario, the expected circuit size can be proportional to the amount of addressable memory.
Static conditions are collapsed at compile time, while loops with statically evaluable termination conditions are automatically unrolled. 

The arithmetic gate language not only supports addition and multiplication operations, but also \textit{wire expansion} and \textit{wire compression} for binary operations. 
Signed numbers are represented as two-complement with a sign bit at the most significant bit, while Boolean expansion is represented as multiple $1$-bit wires. 
We assume that only operations between (signed or unsigned) integers are available. 
Let us consider the following portion of a contract: \textit{"Check if the average salary of the employees is greater than $\$32.5K$"}.
This statement requires a division (by $N$ employees) in order to compute the average value. However, the statement can be converted into the following expression between integers: $\sum_{i=1}^N s_i > 32500N$,
where $s_i$ represents the salary of the $i$-th employee. 

\subsubsection{Pre-processing}
\label{sec_prepro}
A smart contract may consist of multiple files and libraries. The first step of the protocol involves the creation of a single source file containing the full set of instructions required to implement the contract. Header declarations from the header files are imported to the source files, all source files are merged and all pre-processor C directives, macros and constants are evaluated or solved. They include \texttt{\#define} directives and conditional \texttt{\#ifdef} directives.
Moreover, the declaration of the entry point in the source code must have a predefined syntax highlighting input and output types.
The code in Listing 1 shows an example of a C-language contract containing a sum operation between two unsigned integer inputs.

\subsubsection{Creation of the global table of symbols}
A \textit{table of symbols} is a data structure used by a compiler or interpreter to associate each identifier (symbol) in the source code with information relating to its declaration. In this second step, the interpreter detects all the global symbols declared in the source file, e.g.
functions, structures, classes and constants.
For each of these symbols, a hierarchy of local symbols representing the internal declarations of their identifiers is built. At the end of this stage, each global symbol (name, type and value) in the table can be directly addressed for further processing.
One of the global symbols must be the entry point of the contract (i.e. main function \texttt{contract()} in Listing 1). Name, number and type of its parameters are checked against the expected syntax. 

\subsubsection{Line-by-line evaluation}
\label{sec_lineby}
Each source code line is analyzed independently. Local symbols, representing the internal declarations of identifiers, are included in the hierarchy of the global table of symbols. In more detail, this stage is responsible for the following tasks:
(i) \textit{decoding of types}, including the declaration of structures and array, elementary types (Booleans, integers, etc.) and pointers;
(ii) \textit{decoding of expressions}, e.g. unary or binary operations, constants, identifiers, data structures and function calls;
(iii) \textit{evaluation of expressions}, i.e. evaluation of (numeric) expressions which do not depend on the input values;
(iv) \textit{memory allocation}, i.e. temporary storage allocation for the data structures required by the contract functionality.
This stage links all the statements of the arithmetic expression from a spatial (i.e. memory used) and temporal (i.e. operator precedence) point of view. Therefore, each output variable is expressed as a combination of logic and arithmetic operations applied on the input variables.

A generic arithmetic/logic expression is collapsed in order to be represented in an explicit form,
i.e. an arbitrary operator $OP_{i+1}$ is applied to the expression after the operator $OP_i$. The expression is used to create the arithmetic primitives required to represent the contract functionality.

\lstset{aboveskip=4mm, belowskip=0mm, basicstyle={\small\ttfamily}, frame=tb, language=C}
\begin{lstlisting}[caption={Example of C-like contract skeleton.},captionpos=b, float=tp]
struct in_T { unsigned int i1; unsigned int i2; };
struct out_T { unsigned int o; };

void contract(struct in_T *in, struct out_T *out)
{
    unsigned int val = in->i1 + in->i2;
    out->o = val;
}
\end{lstlisting}


\subsection{Generation of the arithmetic primitives}
At this stage, the compiler is ready to make a one-to-one mapping between the operations used to generate the expression and the structures required to implement these functionalities on a circuit.
We denote by $n_{bit}$ the number of bits (\textit{bit-width}) used to represent a signed or unsigned integer in binary. Different computer architectures are characterized by different $n_{bit}$ values. If a client does not know a preferred bit-width value of a worker, its value will be arbitrarily chosen and specified in the header of the circuit.
In the same way compilation is performed for a specific target architecture, knowing the bit-width value may result in a more efficient implementation and execution of the circuit.

\subsubsection{Addition and multiplication operations}
Every arithmetic or Boolean wire $x$ in the circuit can be uniquely identified by a value $id_x$. As for binary variables, we start to count from zero. Addition and multiplication operations are mapped one-to-one into addition and multiplication gates in the circuit. Given two $n$-bit wire inputs, an addition wire output requires $n+1$ bits and a multiplication wire output requires $2n$ bits. For instance, a multiplication between two $n_{bit}$ wires $a$ and $b$ can be represented as:~{\texttt{MUL [$id_a$ $id_b$] TO [$id_c$]}}. 

\subsubsection{Boolean operations}
The full set of Boolean gates can be computed using arithmetic gates, e.g. given two Boolean values $a$ and $b$, we have:
AND$(a, b) = ab$,
OR$(a, b) = 1 - (1 - a)(1 - b)$,
and XOR$(a, b) = (1 - a)b + (1 - b)a$.
All arithmetic operations are performed on $1$-bit width wires.
Bitwise Boolean operations on $n$-bit width inputs require $n$ $1$-bit multiplications (for AND) or additions (for OR). Starting from the least significant output bit, each element is then multiplied by two and added to the next element to build the resulting n-bit integer value.

\subsubsection{Wire expansion}
Wire expansion is usually used to translate an arithmetic wire $a$ to an $n_a$-bit output wire, where $n_a$ is the base-2 logarithm of the maximum value which can be expressed by $a$. For instance, let us consider the following portion of a contract: \textit{"Check if variable [a] is even"}.
Assuming that $n_a = 4$ and $a_0$ represents the least significant bit of $a$, the output of the statement is given by $a_0$. This circuit building block can be expressed as: {\texttt{EXPAND [$id_a$] TO [$id_{a3}$ $id_{a2}$ $id_{a1}$ $id_{a0}$]}}. The compiler may generate only the individual $1$-bit wires used in the rest of the contract, removing the remaining $1$-bit wires, by applying a specific syntax for the optimized wire expander:~{\texttt{EXPAND [$id_a$] TO [$0 \rightarrow id_{a0}$]}}. That is, only the least significant $1$-bit wire (i.e. identifier number zero) is taken, and identifier $id_{a0}$ is assigned to it. The greater $n_a$, the more effective the space optimization\footnote{In this context, we define \textit{space optimization} to be the amount of memory saved to store or transmit the low-level directives used to represent the arithmetic circuit.} may be.

\subsubsection{Negate operation}
The negate operation is necessary to compare two variables, since their difference can be compared to the value zero. Negating an $n_{bit}$-bit wire can be implemented as multiplication by constant $-1$. This constant (on $n_{bit}$ bit) must be represented as: $-1_{n_{bit}} \triangleq  \sum_{i=0}^{n_{bit}-1} 2^i$.

\subsubsection{Equal to zero operation}
This building block for an $n_{bit}$-bit wire $a$ can be implemented as follows: (i) wire expansion on $n_{bit}$ bit $\{a_0, \dots, a_{n_{bit}-1}\}$; (ii) negate each $1$-bit wire (i.e. $a_i \rightarrow b_i$); (iii) multiply the resulting $b_i$ wires: $c = \prod_{i=0}^{n_{bit}-1} b_i$. Therefore, $1$-bit variable $c$ is set to one if and only if $a = 0$.

\subsubsection{Compare to zero operation}
A \textit{greater than} operation can be transformed to a \textit{less than} operation using simple equation substitutions. In the two's complement representation, this operation corresponds to check if the difference between two signed integers is positive or negative (or equal to zero in the case of \textit{less than} or \textit{equal to} operation). The discriminant of the sign of the difference, e.g. $c = a - b$, is given by the most significant bit $x$ in the binary representation: negative numbers are characterized by $x = 1$, while positive numbers are characterized by $x = 0$. This statement can be represented as:~\texttt{EXPAND [$id_c$] TO [$n_{bit} - 1 \rightarrow x$]}.
Depending on the type of comparison (positive vs. negative), the binary value $x$ is required to be negated.

\subsubsection{Conditional statement}
\begin{figure}
\includegraphics[width=.7\linewidth]{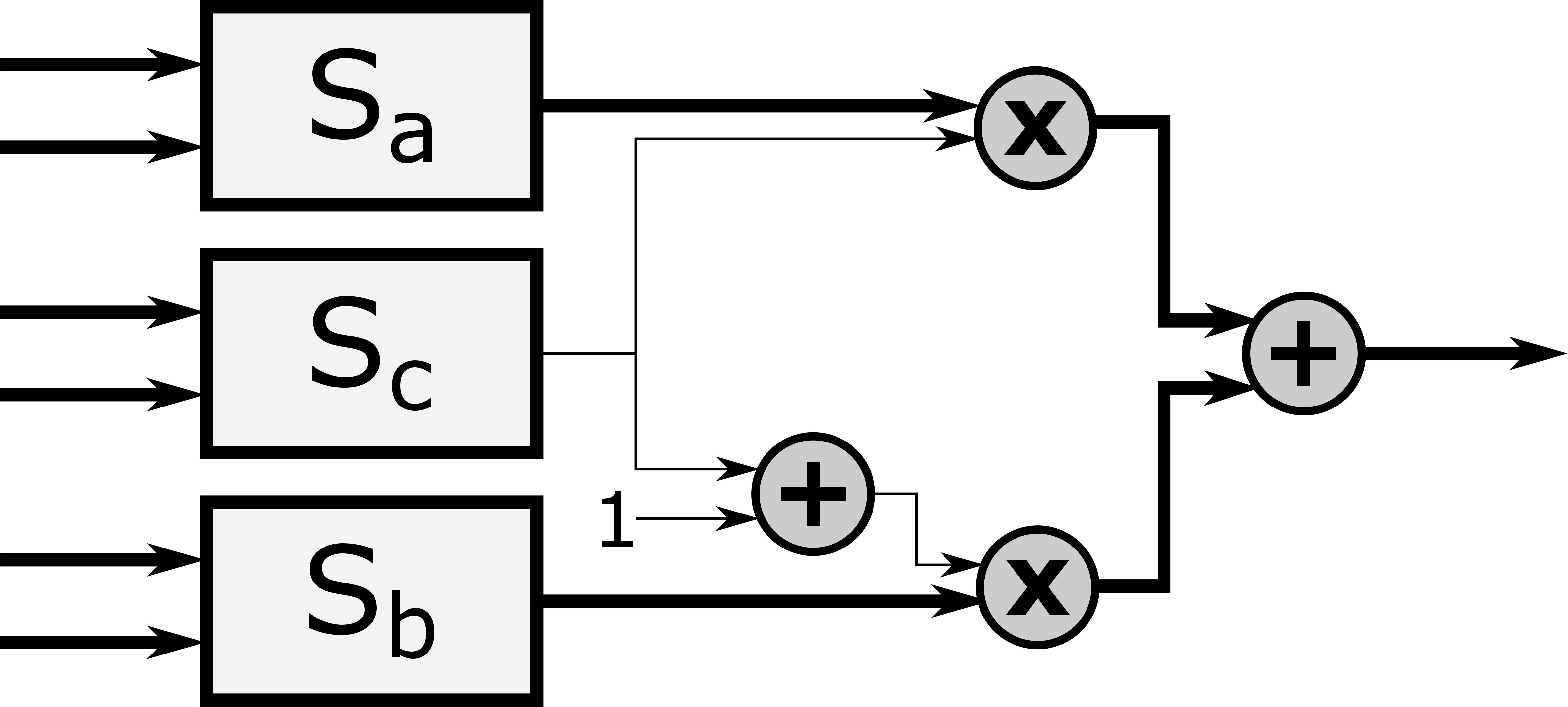}
\caption{Building block implementation of a conditional statement. Depending on the (binary) output of statement $S_c$, statement $S_a$ or statement $S_b$ will be executed. The binary operation $x + 1$ is used to negate $x$.
\label{fig_if}}
\end{figure}
A conditional statement in a high-level language can be expressed in the following form:~\texttt{IF ($S_c$) $S_a$ ELSE $S_b$}. Since the statement $S_c$ depends on the input of the contract, both branches $S_a$ and $S_b$ must be implemented in the circuit. The logic flow is depicted in Figure ~\ref{fig_if}.

\subsubsection{Generation of constants}
Constants' values do not depend on the input wires of the circuit. Using dedicated unary multiplication gates in the form \texttt{mul-by-const-$c$}, we propose the following additional circuitry to generate the constant values required by the contract:
(i) constant \textit{zero} is computed by multiplying an input wire by zero;
(ii) constant \textit{one} is computed by adding one to constant zero and
(iii) any additional constant $c_i$ is computed by using \texttt{mul-by-const-$c_i$} on constant \textit{one}.
Since constants zero and one are always added to the circuit, the implementation of $k$ arbitrary constants requires $k+2$ gates. Constants have a known bit-width as specified by the two's complement standard.


\subsection{Circuit minimization}
The circuit size can be reduced in order to optimize the contract execution time,
the circuit upload/download time and the storage space.
The HLL compiler produces a circuit composed of arithmetic gates. However, complex arithmetic circuits embed \textit{logic submodules} because of conditional and flow control statements. These submodules are still converted to arithmetic circuits, but they are characterized by 1-bit width gate connections.
Since logic and arithmetic 1-bit width circuits are dual, the theory of logic \textit{circuit minimization} can be applied to logic submodules in the arithmetic circuits (see thin wires in Figure~\ref{fig_if}). Moreover, submodules do not share any internal gate, therefore the minimization procedure can be parallelized to reduce the time complexity.

\begin{figure}
\includegraphics[width=\linewidth]{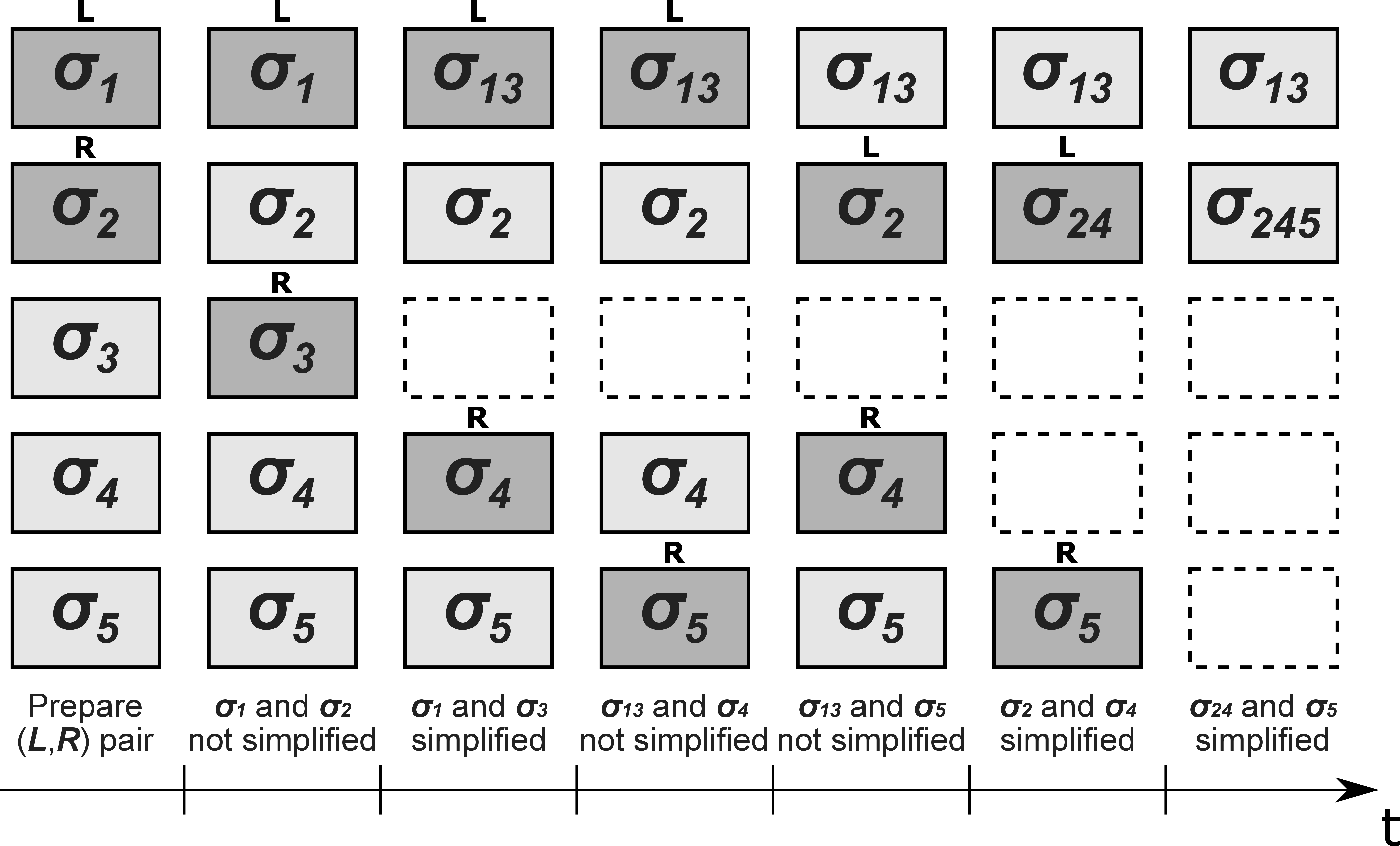}
\caption{Example of prime implicant reduction with $M = 5$. $L$ (left) and $R$ (right) members are highlighted at each step of the process. The auxiliary function $f^{*}$ is reduced to $\sigma_{13} \cdot \sigma_{245}$.
\label{fig_simpl}}
\end{figure}

One of the historical methods used to simplify a Boolean expression $f$ is the Quine-McCluskey algorithm, which returns the complete list of prime implicants\footnote{A product term $\psi$ in a sum of products is an \textit{implicant} of the Boolean function $f$ if $\psi \rightarrow f$. A \textit{prime implicant} of $f$ is an implicant that cannot be covered by implicants with fewer terms. Removing any term from the $\psi$ results in a non-implicant for $f$.} of a Boolean function~\cite{quine}. Moreover, Petrick's method~\cite{petrick} can be used to reduce the number of prime implicants in order to represent $f$ as a composition of \textit{essential prime implicants}, i.e. prime implicants that cover an output of $f$ that no combination of other prime implicants is able to cover. An auxiliary Boolean function $f^{*}$ can be expressed as product of sums $\sigma_i$ of the prime implicants contributing to each output of $f$, i.e. $f^{*}=\prod_{i=1}^M \sigma_i$,
where $M$ is the number of \textit{minterms} (i.e. products) used to express $f$ and terms $\sigma_i$ are represented as follows:
\begin{equation}
\sigma_i=\sum_{k=1}^{|z_i|} z_{ik}.
\end{equation}
Therefore, $z_{ik}$ represents the $k$-th prime implicant contributing to the sum $\sigma_i$.
Starting from $\sigma_1$ and $\sigma_2$, we look for the simplification of the product terms of $f^{*}$ using a set of elementary Boolean rules:
\begin{enumerate}
\item $u (u + v) = u$
\item $u (u^{\prime} + v) = uv$
\item $(u + v) (u + w) = u + vw$
\end{enumerate}
Let us consider the product terms $\sigma_i$ as a list $\{\sigma_1, \dots, \sigma_M\}$.
At each step, two members of the list are compared (as left and right members) and simplified if possible (see Figure~\ref{fig_simpl}). If the first two members $\sigma_1$ (left) and $\sigma_2$ (right) can be simplified, then they will be substituted with a new term $\sigma_{12}$.
In the second step, left member $\sigma_1$ or $\sigma_{12}$ will be checked against right member $\sigma_3$, and the outcome can be $\sigma_1$, $\sigma_{13}$ or $\sigma_{123}$. The third step involves $\sigma_4$ as right member and so on. When all right members are checked, the left member is set to its next member in the list and the right member is set to the next member of the new left member. The process ends when no more checks are due.
The number of simplification steps is bounded from above by $\frac{M(M - 1)}{2}$.
The $u$ term in rules (1), (2) and (3) represents the common part in the (left, right) pair. A cross-check (intersection) between the addends of left and right members is characterized by a time complexity $O(n^2)$ or $O(n\log{n})$ depending on the specific implementation (e.g. na\"{i}ve or sorted lists).

Function $f^{*}$ is covered by each term independently, i.e. $f^{*} = \sigma^{*}_0 + \sigma^{*}_1 + \dots + \sigma^{*}_p$.
Since each term $\sigma^{*}_i$ is actually a product of the set of prime implicants $z_i$, function $f^{*}$ can be covered by the minterm $\sigma^{*}_L$ in $\sigma^{*}$ containing the minimum number of prime implicants, i.e. $f^{*} = \sigma^{*}_L$,
where L = $\argmin_i C(\sigma^{*}_i)$ and function $C(\sigma^{*}_i)$ counts the number of prime implicants $z(\sigma^{*}_i)$ contained in $\sigma^{*}_i$. Finally, the dual function $f$ can be expressed as follows:
\begin{equation}
f = \sum_{k=1}^{C(\sigma_L^{*})} z_k(\sigma_L^*).
\end{equation}

We propose a computationally-optimized heuristic to assign the individual logic submodules to different processing cores for the logic minimization stage, called the LPT greedy algorithm (Longest Processing Time). If the jobs are sorted by their processing time and then assigned to the machine with the earliest end time so far, the scheduler tries to balance the computational load for each transaction.
We assume that the number of initial gates $g_i$ of a logic submodule $S_i$ represents a good estimation of the expected time required to minimize $S_i$. Therefore, given a machine with $N$ cores and $n$ submodules to minimize, submodules $S_i$ are sorted by their $g_i$ value $(1 \le i \le n)$ and then assigned to list $\{list_j\}$ of core $j$ with lowest aggregate $G_j(t^\prime)$ metric at a given time $t^\prime$:
\begin{equation}
G_j (t^\prime) = \sum_{t=0}^{t^\prime} g_i(t)~\forall i \mid S_i \in \{{list}_j\}.
\end{equation}
Alternatively, submodules $S_i$ can be assigned to the processing core in a round-robin fashion without considering the $g_i$ values. Therefore, $S_i$ is assigned to core $j$ if and only if $i$ (modulo $N$) = $j$.

\section {Blockchain interaction}
\label{sec_block}


In validating transactions, the Bitcoin blockchain makes use of the Unspent Transaction Output (UTXO) set. UTXOs consist of two main parts: the amount transferred to the output and the locking script (scriptPubKey) that specifies the conditions to be met in order to spend the output. Bitcoin scripts are written in a bytecode stack-based language called \texttt{Script}. They consist of a sequence of instructions executed linearly, with no jumps backwards. Essentially, to access a specific output, the corresponding locking script must be supplied with parameters that render its result to true~\cite{Zohar:2015}. Standard blockchain scripts can be categorized into five\footnote{Bitcoin (Core) also has Witness versions, e.g. Segregated Witness.} types: Pay-to-Public-Key-Hash (P2PKH), Pay-to-Script-Hash (P2SH), Multisig, Pubkey (P2PK), and null data (\textsf{OP\_RETURN}). In P2SH transactions, the locking script that is replaced by a hash is referred to as the \textit{redeem script}.


In NECTAR,
smart contract verifications are part of the blockchain transaction validation, therefore we store the elements required during the verification stage in the unlocking script.
To illustrate how our solution extends the functionality of the Bitcoin blockchain, we consider a simple smart contract that transfers a token to a certain party, triggered by a condition. For example: \textit{"Transfer Token A (e.g. 100 tulips, 100 shares) from Charlie to Alice on 1st of January 2018 at a cost of 1 BTC that Alice agrees to pay."}
The contractual parties (e.g. Alice and Charlie) agree on the contract conditions over the phone. Charlie creates a transaction (Funding TX) with two inputs and two outputs and signs both inputs with the following:~\textsf{SIGHASH\_SINGLE}~$\vert$~\textsf{SIGHASH\_ANYONECANPAY}.
The first output contains the contract details, while the second output is a P2PKH paying 1 BTC (the price for Token A) to Charlie's address. Charlie does not broadcast this transaction, but he gives it off-line to Alice, who checks that the transaction is correct and the price for Token A is as agreed.

Our proposed protocol makes use of zero-knowledge cryptography in the execution of the contract. Thus, for executing the Funding TX, Alice needs to compute a set of public parameters $(E_K, V_K)$, i.e. the evaluation and verification keys, based on the QAP. The size of $E_K$  depends on the size of the circuit under consideration, i.e. the number of elements making up $E_K$  corresponds to the number of internal multiplication gates of the circuit, while the size of $V_K$ depends on the number of inputs and outputs. 

In the context of smart contracts, it is of general interest to have a public record of the PoC and $V_K$, allowing everyone to verify the validity of the computation and proof. Therefore, Alice needs the services of a worker that uses $E_K$  to generate a PoC verifiable by any third party using $V_K$. For this to happen on the blockchain, we need to make sure that: (i) Alice has provided $(E_K, V_K)$ off-line; (ii) the appropriate $V_K$ is used in the creation and verification of the PoC, and both these parameters are available on the blockchain and can be checked by any party.

As explained in Section~\ref{sec_qap}, the set of keys may consist of a large number of elliptic curve points. This requires the storage of large blocks of data on the blockchain to represent the contract. The challenge, thus, is to arrange the redeem script and the corresponding input script in such a way that they contain the largest amount of data possible.


\subsection{Recording the proof using a P2SH script}

In the current Bitcoin script implementation, it is possible to associate up to 1461 bytes of data with a corresponding input (scriptSig) and a redeem script. This is a well-known means of including larger chunks of data in a transaction. 
Our protocol for storing the public elements in Bitcoin scripts consists of:
\begin{itemize}
\item {Determining how many funding payments to create. This, in turn, depends on the size of the $V_K$.} 
\item{If necessary, split $V_k$ into chunks, i.e.~$V_K=
\{V_{K_1}~\Vert~\dots~\Vert~V_{K_n}\}$,
according to the (maximally) allowed size of data blocks that can be pushed onto the script stack.}
\end{itemize}
\lstset{basicstyle={\small\ttfamily},breaklines,frame=trbl}
The interaction of Alice (client) and Bob (worker) with the blockchain is shown in Figure \ref{fig_main}. Alice chooses a secret value $s$, builds $E_k(s)$ and $V_k(s)$, and creates a locking script that contains the hash of $V_K$, i.e. $\mathtt{H(}V_K\mathtt{)}$.
The locking script is placed inside the funding transaction (Funding TX) and is of the form:
\begin{lstlisting}
OP_HASH160 <H(redeem_script)> OP_EQUAL
\end{lstlisting}
\vspace{2mm}
The redeem script includes the PoC verification. In the case $V_K$ can fit into one block of data, the redeem script is given by:
\begin{lstlisting}
<redeem_script> :=
 OP_HASH160 <H($V_K$)> OP_EQUALVERIFY <PubKey_worker>
 <script_PoC_verification>
\end{lstlisting}
\vspace{2mm}
To allow such construction, a new Bitcoin script opcode that enables pairing verification is required within \texttt{script\_PoC\_verification}.
Here we have omitted the \textit{push-byte} operations, used to push elements onto the stack.  Because of the size of $V_K$ and the constraints of the data pushed onto the stack, the redeem script can also be a succession of hashes of $V_K$, i.e.
\begin{lstlisting}
H( OP_HASH160 <H($V_{K_i}$)> OP_EQUALVERIFY ...
   OP_HASH160 <H($V_{K_n}$)> OP_EQUALVERIFY
   <PubKey_worker> <script_PoC_verification> )
\end{lstlisting}
\vspace{2mm}
where \texttt{H($V_{K_i}$)} is the hash of $V_{K_i}$. The funding transaction is subsequently signed and broadcast by Alice.

When Bob has evaluated the circuit and is able to produce a PoC, he proceeds with creating the corresponding unlocking script. For instance, if $V_k$ fits in a single data block, the script is of the form:
\begin{lstlisting}
<PoC> x y $V_K$ <redeem_script>
\end{lstlisting}
\vspace{2mm}
Here \texttt{x} and \texttt{y} are the inputs and outputs of the circuit, respectively. We thus see that Bob needs to provide the PoC in order to redeem the funds, paid by Alice, for the work.


\section{Conclusions}
\label{sec_conc}
With NECTAR we tackled and successfully solved the problem of improving the privacy, correctness and verifiability of smart contracts that are executed as part of transaction validation on the blockchain. In this paper we showed how any non-specialist programmer can publish a contract on the blockchain without the need to implement any cryptographic protocol. This smart contract is translated under the hood of NECTAR into a set of verifiable equations. Proofs of correct executions are then included in blockchain transactions, demonstrating the practical viability of our solution. 
The ability to support operations like exponentiation or floating point representation will be a significant improvement, giving NECTAR the potential to address complex anonymity concepts and sophisticated financial contracts.

\bibliographystyle{ACM-Reference-Format}


\end{document}